\documentclass[acmsmall,nonacm]{acmart}

\AtBeginDocument{%
  \providecommand\BibTeX{{%
    \normalfont B\kern-0.5em{\scshape i\kern-0.25em b}\kern-0.8em\TeX}}}

\usepackage[switch]{lineno}

\usepackage{booktabs} 
\usepackage{tabularx}
\usepackage{adjustbox}
\usepackage{framed}
\usepackage{enumitem}

\usepackage{ragged2e}
\newcolumntype{P}[1]{>{\RaggedRight\hspace{0pt}}m{#1}}
\usepackage{color, colortbl}
\definecolor{Gray}{gray}{0.94}
\usepackage{pifont}
\usepackage[capitalize]{cleveref}
\usepackage{tcolorbox}
\newcolumntype{L}[1]{>{\RaggedRight\hspace{0pt}}p{#1}}

\usepackage{tikz}
\usetikzlibrary{shapes.geometric, arrows.meta, positioning}

\makeatletter
\def\@copyrightspace{\relax}
\let\@authorsaddresses\@empty
\makeatother
\setcopyright{none}
\renewcommand\footnotetextcopyrightpermission[1]{} 
\author{Sayash Kapoor$^1$, Peter Henderson, Arvind Narayanan
\\
\bigskip
C\MakeLowercase{enter} f\MakeLowercase{or} I\MakeLowercase{nformation} T\MakeLowercase{echnology} P\MakeLowercase{olicy}, P\MakeLowercase{rinceton} U\MakeLowercase{niversity}
\\
\bigskip
\textbf{\today}
}

\begin{document}
\sloppy
\title{\LARGE Promises and pitfalls of artificial intelligence for legal applications}
\renewcommand{\shortauthors}{}
\renewcommand{\shorttitle}{Promises and pitfalls of AI for legal applications}

\begin{abstract}
Is AI set to redefine the legal profession? We argue that this claim is not supported by the current evidence. We dive into AI's increasingly prevalent roles in three types of legal tasks: information processing; tasks involving creativity, reasoning, or judgment; and predictions about the future. We find that the ease of evaluating legal applications varies greatly across legal tasks, based on the ease of identifying correct answers and the observability of information relevant to the task at hand. Tasks that would lead to the most significant changes to the legal profession are also the ones most prone to overoptimism about AI capabilities, as they are harder to evaluate. We make recommendations for better evaluation and deployment of AI in legal contexts.
\end{abstract}

\maketitle

\footnotetext[1]{Author emails: sayashk@princeton.edu, peter.henderson@princeton.edu, arvindn@cs.princeton.edu. Parts of this paper are based on blog posts by two of the authors (\url{https://aisnakeoil.com}).}


\section*{Introduction}

DoNotPay, a U.S.-based AI startup, claimed to sell the services of a `robot lawyer' to help customers prepare legal documents, contest parking tickets and cancel subscriptions~\cite{noauthor_donotpay_2023}. On January 8, 2023, CEO Joshua Browder claimed that the company would pay USD 1 million to any lawyer who used DoNotPay's robot lawyer to argue a U.S. Supreme Court case by using an earpiece to repeat the arguments made by the company's software~\cite{joshua_browder_jbrowder1_donotpay_2023}. Even setting aside the fact that the Supreme Court prohibits electronics in the courtroom, the U.S. has several laws prohibiting the unauthorized practice of law by individuals who are not licensed attorneys. Soon after the announcement, the CEO backed down~\cite{joshua_browder_jbrowder1_good_2023}, and the term `robot lawyer' was changed to `AI consumer champion' on the company's website~\cite{noauthor_donotpay_nodate}. Still, the company is facing multiple class-action lawsuits~\cite{noauthor_analysis_nodate}. 

This was far from the first time when technology was claimed to replace a lawyer, and it will not be the last. 
After all, 
the company had been claiming to sell the services of a robot lawyer for more than four years. Claims about lawyers being replaced by digital technology predate the company. A 2011 New York Times headline read: `Armies of Expensive Lawyers, Replaced by Cheaper Software.'~\cite{markoff_armies_2011} Since the article was published, the number of lawyers in the U.S. has actually \textit{increased} by eight percent~\cite{statista_research_department_us_2023}. How do we separate true advances from hype?

In this position paper, we argue that the kinds of legal applications we can legitimately use AI for should be determined by the evaluations that reflect said use in the real world. 
It is easy to get caught up in the hype, particularly for impressive demonstrations of generative AI that can be used to create text, images, or other forms of media and is often trained on a vast amount of existing data. Many recent instances of AI that have received widespread attention are examples of generative AI~\cite{noauthor_introducing_nodate,noauthor_dalle_nodate,noauthor_make--video_nodate}.
In the law, some of this attention has focused on claims of text-based language models' improvements in legal reasoning ability, such as OpenAI's claims that GPT-4 can pass the bar exam. Yet, this is not evidence that GPT-4 is becoming as capable as lawyers: after all, it is not a lawyer's job to answer bar exam questions all day. While generative AI is our main focus, we also look at AI for predicting the outcomes of court cases and making decisions about people (such as AI used for predicting a defendant's risk of recidivism). 

The types of legal AI we analyze roughly correspond to the types of AI outlined in \citeauthor{diver_typology_2022}'s typology of legal applications~\cite{diver_typology_2022}, though at a coarser level of granularity. We analyze three broad uses of AI in the legal domain: (i) tasks involving information processing, such as summarization or legal information retrieval; (ii) tasks involving creativity, reasoning, or judgment, such as preparing legal filings; and (iii) tasks involving predictions about the future, such as criminal risk prediction as well as predicting the outcomes of court decisions. 
Of course, the lines separating these applications are blurry, but the high-level categories can offer useful insights about how AI applications should be evaluated and how useful they can be in the real world. 

These applications vary in how difficult they are to evaluate.
For some, evaluation is relatively easy.
For example, a tool that categorizes a request for legal advice into particular areas of law (an example of an information-processing task) can be evaluated by comparing against corresponding labels from lawyers performing the same task.\footnote{As in, for example, the Learned Hands project.~\cite{stanford_legal_design_lab_learned_2018}}
In contrast, there is no clear `correct' answer for other types of AI. For instance, if generative AI is used to prepare a legal filing (an example of a task involving creativity, reasoning, or judgment), there is no single correct answer on how the document should be written---reasonable people can disagree on what strategies to take. Tasks that are harder to evaluate also tend to be those that would lead to the most significant changes in the legal profession. If AI could be useful for consequential legal tasks like preparing legal filings, that would have much broader implications for the future of legal professionals compared to labelling text for different areas of law.

In our analysis, we look at the challenges that arise in meaningful evaluations of AI in legal settings and offer recommendations for overcoming them. We argue that evaluations should be used to identify how well AI does on a given task and which types of tasks AI can be useful for~\cite{diver_typology_2022}.

\section*{Information processing}

Many legal tasks involve processing information. Examples include summarizing court cases or long legal documents, translating text from one language to another, redacting sensitive information from documents before broader release, e-discovery to find relevant documents for litigation and legal information retrieval. 

With the widespread adoption of generative AI, there have been many claims that it will revolutionize legal information processing. Compared to the other types of legal tasks we look at in the next two sections, evaluating information processing tasks is more straightforward. This is because:
\begin{itemize}
    \item There is generally a \textit{clear correct answer}: given information about the features used in the model and the model's output, it is easy to determine if the model's output is correct, such as in the task of categorizing legal requests by area of law~\cite{stanford_legal_design_lab_learned_2018}.\footnote{There are, of course, some exceptions where evaluation is more ambiguous even within information processing, but the majority of cases in this category will be more straightforward to evaluate.}
    \item There is \textit{high observability} of the features relevant for decision making: The features relevant for using AI for information-processing tasks are available as inputs to the AI system.
\end{itemize}

These factors make it easier to develop valid evaluations for AI used for information processing. As a result, generative AI for information-processing tasks can be deployed based on evidence and robust evaluations. Still, claims about generative AI being a revolution might be overstated, and several nuances make a blanket assessment of generative AI for information processing hard.

\paragraph{For legal experts, generative AI for information processing is an evolution, not a revolution.} A major reason why chatbots are exciting to the general public is that they can be instructed in natural language to perform tasks for which software may not have previously existed. But for tasks for which natural-language processing software did exist, the advent of large language models has generally led to an evolutionary improvement in accuracy. In law, software for information-processing tasks is nothing new. Automated tools for legal summarization have existed for over a decade~\cite{markoff_armies_2011}. The same goes for many other information processing tasks, with entire companies built on the promise of automating legal information processing dating back decades. Recent instruction-tuned language models (chatbots) cannot necessarily outperform models fine-tuned on law-specific datasets~\cite{chalkidis_chatgpt_2023}. Further, many information-processing tasks can also be carried out by professionals without a law degree. For these reasons, while large language models offer improvements over existing tools --- possibly in terms of accuracy but especially in terms of cost, by decreasing the amount of task-specific software development required --- they do not drastically change legal information processing for experts.  

\paragraph{We need to better understand how generative AI impacts laypeople.} The ability of chatbots to follow natural language directions means laypeople can use them to perform information-processing tasks, such as translation or getting pointers to relevant legal rules. Everyday users have increasingly turned to technology for legal advice in the past---for instance, a 2019 survey in the U.S. found that more people used the internet (31\%) than lawyers (29\%) for legal advice, and that 63\% of the people surveyed used information they found on the internet as a factor to resolve their legal problems~\cite{gramatikov_justice_2021}.

Yet, there is a paucity of evidence about how chatbots affect users who turn to them for information-processing tasks such as legal information retrieval or translation. Understanding how well they work is hard without naturalistic evaluations of everyday users who use chatbots. Errors in the outputs of chatbots on such tasks can be catastrophic. For example, asylum applications for refugees can be rejected if machine translation introduces errors because they cannot accurately infer context~\cite{deck_ai_2023}. It is unclear how people are using generative AI for such tasks. Research should help inform best practices for the use of AI by laypeople.

\paragraph{Unresolved limitations make the adoption of language models challenging.} Language models for information processing suffer several unresolved issues that may pose challenges in shifting from existing solutions to language-model-based ones. A key limitation is their propensity to output incorrect information, often known as hallucinations~\cite{zhao_reducing_2020}. This is a significant hurdle in their adoption in consequential legal settings. While there are many ongoing efforts to improve factual accuracy~\cite{shuster_retrieval_2021}, it is as yet an unsolved research problem. As a result, the outputs of language models must be closely verified before they can be used in consequential settings.

\paragraph{Some information-processing tasks are harder to evaluate than others.} Even within information-processing tasks, ease of evaluation is a spectrum. 
For categorizing cases by area of law, legal experts can label the correct answer~\cite{stanford_legal_design_lab_learned_2018}, but in cases where there might be multiple areas of law implicated, labels might have higher rates of disagreement by experts.
Similarly, for tasks involving transcription or redaction, it is sometimes easy to create a clear source of ground truth based on past data. Yet, in adversarial settings, lawyers might disagree on how much context to redact and litigate over the issues. In \textit{Kaiser Aluminum Warrick, LLC v. US Magnesium LLC},\footnote{WL 2482933 (S.D.N.Y. Feb. 27, 2023)} for example, parties disputed how much information should be redacted in documents produced during discovery and ultimately the court ordered the producing party to unredact information that was relevant to the case.
For translation, evaluations must account for inherent ambiguity---such as when a source language uses gendered terms and a target language does not, there is a lack of context to disambiguate a term, or an idiom does not have a clear direct translation.
And parties dispute how e-discovery systems are evaluated, with requesting parties generally seeking to discover more information and producing parties wanting to reveal less~\citep{guha_vulnerabilities_2022}.
Nonetheless, such disagreements are generally over atypical cases, and the bulk of information processing tasks will have consensus answers when polling a larger pool of annotators.

\section*{Creativity, reasoning, or judgment}

Several legal tasks involve creativity, reasoning, or judgment. They range from tasks involving writing, such as preparing drafts of legal filings, to tasks involving judgment, such as automated mediation and dispute resolution. These tasks typically involve significant expertise and labour to get right. 
In contrast to information processing, if AI could indeed automate such tasks, the impact on the legal profession might be huge. When OpenAI announced its GPT-4 language model, it claimed the model could pass a `simulated bar exam with a score around the top 10\% of test takers'~\cite{martinez_re-evaluating_2023}. This led to much speculation about whether AI would soon replace lawyers, presumably because the tool could perform tasks requiring expertise and creativity. 

But what does a high score on the bar exam mean---and more generally, how much can we trust benchmark evaluations? Here, we outline several concerns underlying evaluations of language models in legal settings that make it hard to trust their applicability to real-world legal tasks. We then provide recommendations for improving evaluations and outline tasks for which AI can be evaluated well and is arguably underutilized.

\subsection*{Hurdles in evaluating language models}

\subsubsection*{Contamination}
Contamination refers to including the same data in the training and evaluation data sets for a model~\cite{magar_data_2022}. This can lead to overoptimistic estimates of model performance since a model can simply memorize solutions in its training set instead of being able to answer new questions. It is possible that evaluations such as OpenAI's claims about bar exam performance are overoptimistic due to contamination, but it is hard to know for sure due to the training and fine tuning data being proprietary.

However, as an illustration of the plausibility and seriousness of contamination, consider a different benchmark that OpenAI evaluated GPT-4 on. To benchmark its coding ability, OpenAI evaluated it on problems from Codeforces, a website that hosts coding competitions. The training data cutoff for the original GPT-4 model was September 2021. The model could correctly answer most Codeforces questions from before its training date cutoff, but could not answer questions after its training date cutoff correctly~\cite{horace_he_chhillee_i_2023}. This strongly suggests that the model memorized solutions from its training set---or at least partly memorized them, enough to fill in what it couldn’t recall. That is, instead of developing the capability to answer \textit{new} coding questions, it could only answer questions it had already been trained on.
(The Codeforces results in the paper were not affected by this, as OpenAI used problems from recent Codeforces competitions, resulting in the model being evaluated on fresh problems not in the training set. Sure enough, GPT-4 performed very poorly~\cite{openai_gpt-4_2023}.) 

To be clear, a temporal discontinuity in benchmark performance, such as in the case of GPT-4's performance on Codeforces, strongly implies contamination, but the lack of such a discontinuity does not imply the opposite. Without access to the data used to train and fine tune a model, researchers can only make informed guesses about the absence of contamination since there is no guarantee that a model is not already trained on later versions of a benchmark. For example, OpenAI could fine tune GPT-4 on more recent versions of the bar exam (even inadvertently) if a user inputs exam questions into ChatGPT.

\subsubsection*{Lack of construct validity}\label{subsubsec:genai_eval_construct-validity} Construct validity refers to the extent to which an evaluation accurately represents and measures the construct it is designed to assess. For the bar exam, the construct might be the extent to which a lawyer has the necessary preparation to serve clients effectively. The assumption is that humans taking exams generalize the skills tested by the exam to a wider range of relevant tasks. 

Unfortunately, it is well known that bar exam questions are not representative of the tasks professionals do in the real world---something that critics of the bar exam regularly lament, resulting in recent restructuring of the bar exam~\citep{Sloan2023} and proposals for alternative pathways to certification based on real-world training~\citep{ChingHershkowitz2023}.
Specifically, the bar exam overemphasizes subject-matter knowledge and underemphasizes real-world skills, which are far harder to measure in a standardized, computer-administered way. In other words, not only does it emphasize the wrong thing, it overemphasizes precisely the thing that language models are good at.

Memorization is a spectrum. Even if a language model has not seen an exact problem on a training set, it has inevitably seen examples that are pretty close, simply because of the size of the training corpus. That means it can get away with a much shallower level of reasoning. This issue is also referred to as \textit{task contamination}~\cite{li_task_2023}. As a result, legal benchmarks don’t necessarily give us evidence that language models are acquiring the kind of in-depth reasoning skills that human test-takers might have. While this inference might already be somewhat dubious for humans, it is unfounded for language models that might take all sorts of shortcuts~\cite{geirhos_shortcut_2020} and memorize key information to come to the right answer without generalizing in any way.

In some real-world tasks, shallow reasoning may be sufficient---for example, it could be enough to build a chatbot to help applicants prepare for the bar exam where similar scenarios have played out thousands of times in textbooks and court cases. But the world is constantly changing, so if a bot is asked to analyze the legal consequences of a new fact pattern in the context of new judicial decisions, it does not have much to draw upon. In short, tests designed for humans lack construct validity when applied to bots.

Benchmarks are already wildly overused in AI for comparing different models~\cite{raji_ai_2021}. They have been heavily criticized for collapsing a multidimensional evaluation into a single number~\cite{thomas_reliance_2022}. As we've discussed, using benchmarks to compare humans to AI introduces a further set of problems. So if an AI developer's goal is to predict how well it will do on real-world legal tasks, measuring bar exam performance is simply not a suitable approach.

\subsubsection*{Prompt sensitivity}
Another issue with evaluating language models is their sensitivity to the user's prompts. Small changes to the prompt can significantly impact the model's outputs~\cite{guha_legalbench_2023}. So it is important to understand how language models are used in order to construct valid evaluations.
Unfortunately, we are entirely in the dark about how users use these models in the real world. Since model developers do not share information about model use, we currently have few ways to study many important questions about language models. 

Prompt sensitivity leads to several challenges in AI evaluation. First, in programmatic use, where developers are writing prompts for legal applications (instead of legal professionals or end users directly using a language model), performance could improve with better prompting, so measured results provide a lower bound of how well the tools work. In some cases, performance could also degrade as the model's behaviour changes over time~\cite{chen_how_2023,narayanan_is_2023}. 
Second, in use by legal professionals, prompt sensitivity means that results are conditional on users being trained on proper prompting techniques. Recent large-scale evaluations of language model performance start to expand the scope of evaluations on a wider range of legal tasks~\citep{guha_legalbench_2023}, but even in these cases, benchmark creators pick a fixed set of prompts that are used across evaluations. It is possible that a user, particularly those not knowledgeable enough about the legal domain or the limitations of language models, could see drastically different performance on the same tasks if they do not craft their prompt in the same way as the evaluation benchmark. Even ordering few-shot examples in a prompt differently can affect performance by double-digit percentage points~\citep{lu2021fantastically}.
Third, in use by non-professionals, the state of evaluation is even worse. The lack of naturalistic datasets means that we cannot evaluate how often chatbots respond to legal questions with useful answers as opposed to irrelevant or inaccurate ones since we do not know how everyday users interact with these models in the real world.

\subsection*{Recommendations for developers of legal AI}

\subsubsection*{Improve construct validity by involving legal experts in evaluation} 
Many current evaluations of LLMs are general purpose: they measure the efficacy of language models on general tasks such as summarization, retrieval, or factuality. However, these evaluations do not tell us much about how LLMs can aid legal professionals in their day-to-day tasks. The involvement of legal experts in designing and conducting evaluations is necessary to improve the status quo. Without their involvement, benchmarks for testing language models on legal tasks will likely suffer from construct validity issues.

Such evaluations can be both quantitative and qualitative. An interdisciplinary group of lawyers and AI experts created the LegalBench benchmark for evaluating language models on various legal reasoning tasks~\cite{guha_legalbench_2023}. This is an example of a quantitative evaluation created by professionals to measure the usefulness of generative AI in their profession.
But there are reasons to think that qualitative studies of professionals and how they could use AI are likely to be even more useful, since these tools are so new that we still need consensus on what the right questions to ask are. To our knowledge, such qualitative studies have not yet been conducted for legal professionals. However, in other professions, notably medicine, several such studies have been conducted, which can inform such evaluations in the law~\cite{noy_experimental_2023,nayak_comparison_2023,abouammoh_exploring_2023}.

\subsubsection*{Develop naturalistic evaluation methods}
As outlined in our discussion of prompt sensitivity, a major limiting factor in current evaluations of language models is the lack of transparency around how users actually use these models on a day-to-day basis~\cite{bommasani_foundation_2023, narayanan_generative_2023}. 
Without knowing how users interact with LLMs, it is hard to understand what limitations must be addressed and how evaluations can best be constructed to represent typical use cases.
To improve the construct validity of current evaluations and prevent evaluations from falling prey to prompt sensitivity, researchers can conduct naturalistic evaluations of people using LLMs that closely model their use in the real world. For example, \citet{zheng_lmsys-chat-1m_2023} released a dataset of user conversations with 25 different LLMs over three months. Similar datasets collecting real-world interactions with users asking legal questions would improve our understanding of how users use LLMs for legal tasks and, in turn, improve evaluations.

\begin{figure}
\centering
\includegraphics[width=\textwidth]{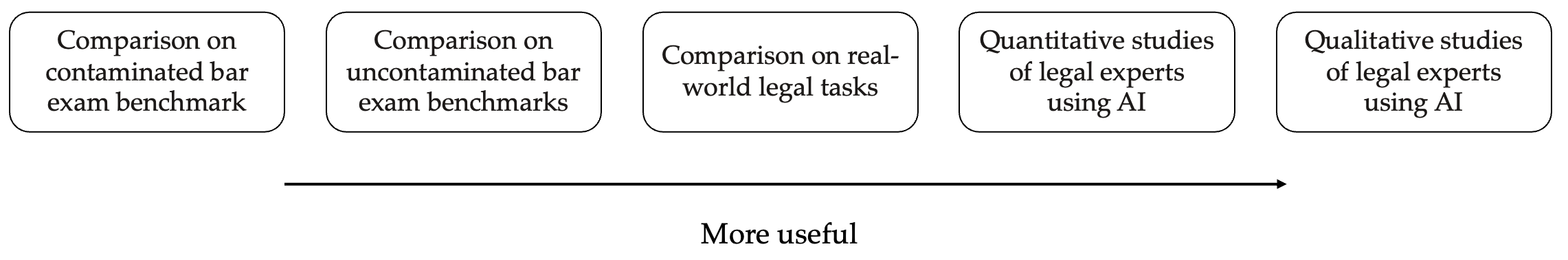} 
\caption{\textbf{Types of evaluations of generative AI}. Current evaluations of AI are often based on exam benchmarks meant for humans, such as the bar exam, and suffer from contamination: overlaps between the training and evaluation datasets. Comparing the performance of these models on real-world tasks, especially those curated by legal experts, is more likely to be useful. Since the use of generative AI is nascent, qualitative studies that observe how legal experts use these tools for day-to-day tasks are likely to be a more useful, if expensive, way of evaluating these tools.}
\label{fig:eval-types}
\end{figure}

\subsubsection*{Communicate the limitations of current LLMs}\label{subsubsec:genai_improving_communication} Recent cases of lawyers misusing language models have made the headlines~\cite{weiser_heres_2023, wagner_this_2023}. Language models can fabricate information even while presenting it authoritatively~\cite{zhao_reducing_2020}. When users are unaware of these limitations, it could result in massive professional damage. Several lawyers have been sanctioned for fabricating information in legal filings. Even when a language model is trained on accurate text, such as a filtered dataset of past legal documents, it is not guaranteed to produce accurate outputs~\cite{michael_c_dorf_law-specific_2023}. These cases highlight the need for better communication of these limitations for end users by companies providing these services~\cite{vincent_openai_2023}. 
Developers have added some disclaimers to language models to reduce such errors. For example, OpenAI says, `ChatGPT can make mistakes. Consider checking important information' in a small font at the bottom of the ChatGPT chatbox. Anthropic goes one step further. Its disclaimer is more clear about the limitations (`Claude is in beta release and may display incorrect or harmful information,'). When the output contains URLs, there is also a disclaimer about links potentially being inaccurate.
Some judges have also issued chambers' rules to clarify how lawyers should explicitly account for their use of AI~\cite{hon_bernice_bouie_donald_generative_2022}. 

\subsubsection*{Use AI in narrow settings with well-defined outcomes and high observability of evidence.} 
\label{subsubsec:AInarrow}
In a more constrained, highly issue-specific, low-stakes setting, it may be possible to construct a thorough evaluation for AI confidence. 
One type of application that meets these criteria is checking errors in various legal documents and filings~\citet{bommasani2021opportunities}.
The Social Security Administration already uses a simple model to spot issues with judgments that might lead to a remand of the judgment on appeal~\citep{glaze2021artificial}. One mistake flagged by the automated system is when the adjudicator's opinion does not address a medical claim made in a benefits claim in their denial of benefits judgment~\citep{glaze2021artificial}. Such a mistake would almost certainly result in a remand of the decision on appeal. 
To make such an assessment, the system does not need any additional information beyond the benefits claim and the text of the decision. That is, the system operates under full observability, allowing thorough evaluations to be conducted.

Similar technology could be developed and deployed in a wide range of settings where easy-to-spot errors in initial filings are prevalent. In particular, over 86\% of patent applications received at least one non-final rejection~\citep{carley2015probability}, so semi-automated checks for common errors could reduce costs to both the filing party and the United States Patent and Trademark Office. Nonetheless, even in these cases, automated judgments should be made with extreme caution. Deployments should be structured to favour helpful, informative recommendations to both parties in a dispute rather than being used as a binding mechanism. And a thorough appeals process should be available.

These tasks are distinct from the more general case of using AI to predict court case outcomes, a more problematic application that we discuss in the next section.  
First, they are constrained to a single or small handful of issues, which makes it possible to sample sufficient data to cover typical use cases.
Second, the model has (or should ideally have) access to the same information as the adjudicator. This is typically not true of general-purpose judgment prediction tasks (\cref{sec:predcourt}).

\section*{AI for making predictions about the future}\label{sec:preds}

In recent years, over a hundred research papers have claimed to predict court outcomes using AI based on text from court proceedings. Such predictive abilities could be useful to lawyers in guiding legal strategy or businesses to assess potential litigation risks. AI has also been used to make consequential decisions about people, most notably pre-trial detention and parole in criminal justice. 

In this section, we identify shortcomings that plague these applications and question the use of predictions in legal settings.

\subsection*{Predicting the outcomes of court decisions}\label{sec:predcourt}

\citet{medvedeva_legal_2023} systematically review 171 papers claiming to predict court decisions. They find severe shortcomings in the literature they review. Their main finding is that the vast majority of papers claiming to predict the outcomes of court judgments do not try to solve this problem at all. In many cases, the papers solve a related but ultimately less helpful problem: they use the judgment text containing the final judgment to `predict' the verdict. Since the text of the final judgment includes the verdict, these studies do not provide real-world evidence of the usefulness of AI in judgment prediction. In sum, only 12 of 171 papers (7\%) end up predicting court decisions.

This study follows a smaller-scale study to evaluate predictions of court decisions, where \citet{medvedeva_rethinking_2023} point out that such errors could be caused by insufficient knowledge of the datasets being used in judgment classification and inadequate steps taken to filter out information about the verdict from the dataset. This highlights the need for both legal and AI expertise for useful applications of AI in legal settings.
Moreover, for the small minority of papers that actually predict court outcomes, the accuracy of the resulting models is much lower. 

The low accuracy demonstrates that automating judgments from the text of legal cases is hard. This is not surprising: legal outcomes depend on the context and specifics of cases, the available documents might not comprise the entirety of the context of the case being adjudicated, and the specific judgment might depend on a specific judge's (or set of judges') interpretation of the arguments. In addition, there is significant variability across different jurisdictions, meaning the amount of data that can be used to train AI to automate judgments in any specific jurisdiction is small. Finally, the judgments made over time evolve with changes to the specific judges, the set of past cases comprising precedent, legislation and many other factors.

\citeauthor{medvedeva_legal_2023}'s findings also point to the problem of contamination. Since the text of the judgment also contains the verdict, the model essentially has access to the answers while making predictions---like teaching to the test, this vastly inflates the accuracy of the resulting models, leading to exaggerated performance estimates. 
This is a well-known issue in machine learning. In traditional machine learning research, it is called \textit{data leakage} or simply \textit{leakage}, and it affects hundreds of papers across scientific fields. While there are no perfect solutions for fixing leakage, there are several steps researchers can take to prevent leakage in their models~\cite{kapoor_leakage_2023}.

This does not even begin to address the potential for biases, sensitivity to inputs and other challenges for evaluating legal judgment prediction tasks. The challenges with evaluation should limit where and how judgment prediction tasks are used. A well-evaluated judgment prediction system could be used to better understand what properties of briefs could lead to poor outcomes (e.g., finding common errors). This would serve as a suggestion to attorneys that might miss common errors but not result in any binding outcome and could be ignored by the attorney.
On the other hand, such systems should not be used to make decisions or final recommendations to a decision-maker.

\subsection*{Predictive AI for making decisions}\label{sec:predai}

In addition to research claiming to predict court outcomes using AI, AI-based predictions are also used to make decisions about people. We call such applications predictive AI. Predictive AI suffers from pervasive shortcomings that may nullify the claimed benefits. A closer account of these shortcomings can help us understand why such systems fail.

\paragraph{Low accuracy of deployed applications.} 
A common application of predictive tools in criminal justice is to predict recidivism. A 2016 ProPublica investigation found that COMPAS, a widely used algorithm to predict the risk of recidivism for defendants, had twice as many false positives for Black defendants as White ones~\cite{angwin_machine_2016}. Perhaps more surprisingly, the investigation found that the overall accuracy of the algorithm was only around 65\%. In a follow-up study, \citet{dressel_accuracy_2018} found that this accuracy was no more accurate than predictions made by people without any background in criminal justice. 
Notably, the majority of defendants predicted to be at high risk of committing violent crimes do not go on to recidivate. 
These simple models distil into these few features a model of a person's entire future life for the next few years.
They have no access to private information, like a defendant's mental state or intentions, nor can they model defendants' attempts to seek help. 

Another problem is \textit{distribution shift}: when the data used to train an ML model differs from the population on which the model is eventually deployed, models are unable to adapt well. A machine learning tool called Public Safety Assessment (PSA) is used in U.S. courts in over half the states. Like COMPAS, if the tool predicts that a defendant has a high risk of re-offending, bail could be denied. PSA is trained on data from 1.5 million cases across the country. But crime patterns in specific regions differ from nationwide averages in important ways, which means that it fails catastrophically in some areas. \citet{corey_how_2019} highlights that in Cook County, Illinois, the rate of violent recidivism is \textit{ten times} lower than the nationwide data that was used for training PSA. 
Distribution shift is an open research problem in machine learning~\cite{geirhos_shortcut_2020}, and affects most predictive AI applications where the population of interest differs from training data~\cite{wang_against_2023}. 
 
Where dynamics are known and stable over time, and information is readily available, prediction is possible---as in physical sciences, where we can build reliable approximations of aspects of the world that we are modelling. Yet this is not true of predictive AI in law, where fundamentally, most predictions will be about people and societies.

Predictive AI has even more limitations in practice. 
Vendors sell predictive AI based on the promise of full automation and elimination of jobs, but when it performs poorly, they retreat to the fine print, which says that the tool should not be used on its own. The responsibility for ensuring that a predictive AI system works well is spread thinly across multiple stakeholders, often deliberately~\cite{paris_martineau_toronto_2022}. 
The individual decisions made by these systems also tend not to be contestable by decision subjects, as vendors claim that the logic of the tool is a trade secret~\cite{jackson_setting_2020,moore_trade_2017}.
In most cases, decision makers (such as court systems) do not develop predictive tools in house---tools that might be tailored to their specific needs and those of the populations that they serve. Instead, they purchase or license one-size-fits-all products from AI vendors. This exacerbates issues with evaluations since the decision subjects or civil rights advocates cannot easily push back against vendors' claims. 

These issues are not specific to the examples we list above. In an analysis of eight predictive AI applications across domains, \citet{wang_against_2023} found that these issues are widespread in domains such as finance, insurance, child welfare and medicine, in addition to criminal justice.
Given the propensity of such applications to failure, predictive AI in the legal domain needs to be held to a much higher standard to ensure that it functions as its developers claim. This requires much stronger transparency by the developers, clear mechanisms to ensure contestability to decision subjects, and evaluations that go beyond just the technical specifications of these tools into the societal impact of these tools. 

The use of AI for prediction, whether court decisions or recidivism, fundamentally differs from information processing tasks and tasks involving creativity, reasoning, or judgment. They attempt to predict the future without sufficient observability of relevant features and lack data to form a robust model of the world that would allow for accurate predictions. Instead, they rely on extremely rough generalizations and approximations using simple linear models (when the underlying dynamics are far from linear).

\section*{Conclusion}

\begin{figure}
\centering
\includegraphics[width=\textwidth]{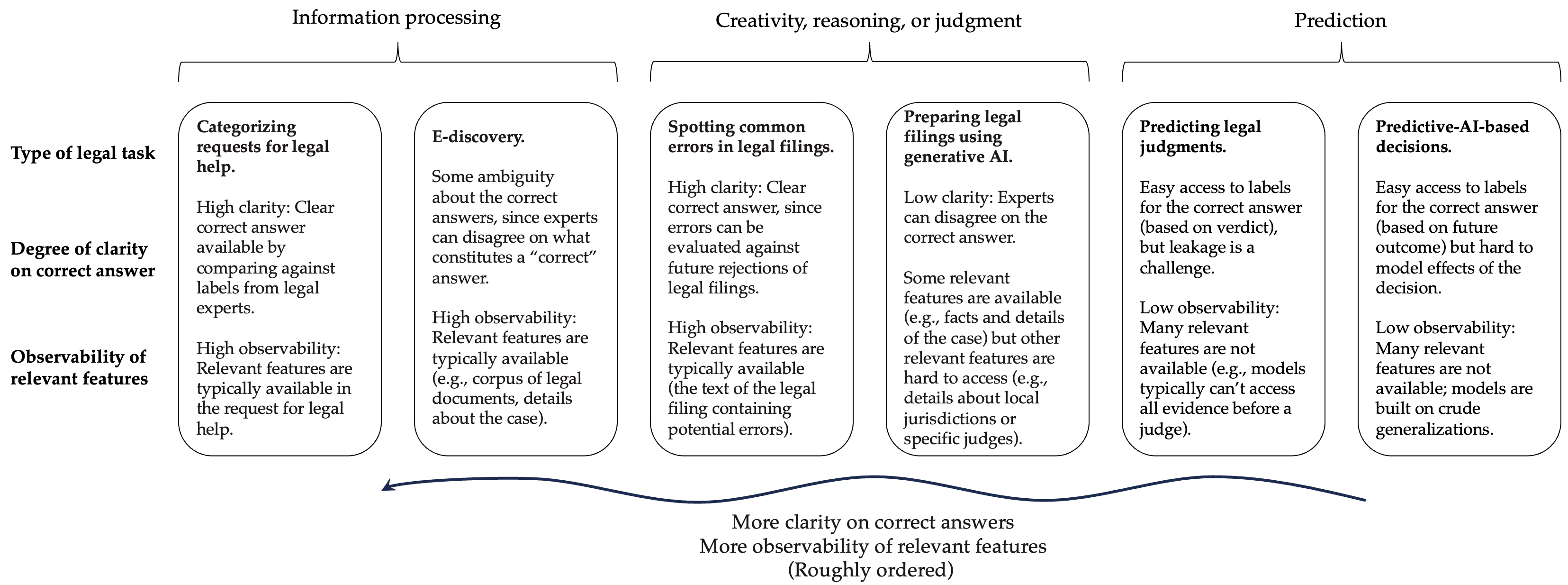} 
\caption{\textbf{Variation in the difficulty of evaluating AI for legal tasks}. We categorize difficulty along two dimensions: clarity on correct labels and observability of relevant features.
Some tasks, such as AI for categorizing requests for legal help by area of law, have clear correct answers, whereas for other tasks, such as preparing legal filings using AI, there is no clear right answer, which makes evaluation hard. 
Similarly, for some applications, all relevant features are available, such as for spotting common errors in legal filings. For others, relevant features are not (or cannot be) available, such as for predictive AI. 
As we proceed from right to left, the clarity of correct answers and observability of relevant features \textit{roughly} increases.}
\label{fig:eval-difficulty}
\end{figure}

The effective deployment of AI in legal contexts requires shifting from technical evaluations to robust socio-technical assessments carried out in the specific context in which an AI system would be deployed. While past machine learning applications did not consider such evaluations because they were cost prohibitive, this change is necessary due to the complex nature and societal impact of AI applications in the legal field. \cref{fig:eval-difficulty} illustrates how the difficulty in evaluating legal applications of AI varies over the three types of tasks we discussed.

To answer the question `What can I use an AI system for?', it is essential first to answer `How was this AI system evaluated?'. Unfortunately, the current state of AI evaluations leaves much to be desired. 

\textbf{Acknowledgments.} We thank Mireille Hildebrandt and the attendees of the Cross-disciplinary Research in Computational Law (CRCL 2023) conference for their feedback. 
We also thank attendees of the World Intellectual Property Organization Judges Forum and the Harvard Journal of Law and Technology speaker series for their feedback on talks based on this paper.
We are grateful to Angelina Wang, Emily Cantrell, Matthew J. Salganik and Solon Barocas for the conversations and collaborations that informed this paper. 
%


\bibliographystyle{plainnat}
\bibliography{references,refs2}

\begin{thebibliography}{56}
\providecommand{\natexlab}[1]{#1}
\providecommand{\url}[1]{\texttt{#1}}
\expandafter\ifx\csname urlstyle\endcsname\relax
  \providecommand{\doi}[1]{doi: #1}\else
  \providecommand{\doi}{doi: \begingroup \urlstyle{rm}\Url}\fi

\bibitem[noa({\natexlab{a}})]{noauthor_analysis_nodate}
{ANALYSIS}: {DoNotPay} {Lawsuits}: {A} {Setback} for {Justice} {Initiatives}?, {\natexlab{a}}.
\newblock URL \url{https://news.bloomberglaw.com/bloomberg-law-analysis/analysis-donotpay-lawsuits-a-setback-for-justice-initiatives}.

\bibitem[noa({\natexlab{b}})]{noauthor_dalle_nodate}
{DALL}·{E} 3, {\natexlab{b}}.
\newblock URL \url{https://openai.com/dall-e-3}.

\bibitem[noa({\natexlab{c}})]{noauthor_donotpay_nodate}
{DoNotPay} - {Your} {AI} {Consumer} {Champion}, {\natexlab{c}}.
\newblock URL \url{https://web.archive.org/web/20230730013643/https://donotpay.com/}.

\bibitem[noa({\natexlab{d}})]{noauthor_introducing_nodate}
Introducing {Claude}, {\natexlab{d}}.
\newblock URL \url{https://www.anthropic.com/index/introducing-claude}.

\bibitem[noa({\natexlab{e}})]{noauthor_make--video_nodate}
Make-{A}-{Video}, {\natexlab{e}}.
\newblock URL \url{https://makeavideo.studio/}.

\bibitem[noa(2023)]{noauthor_donotpay_2023}
{DoNotPay} - {The} {World}'s {First} {Robot} {Lawyer}, January 2023.
\newblock URL \url{https://web.archive.org/web/20230101170502/https://donotpay.com/}.

\bibitem[Abouammoh et~al.(2023)Abouammoh, Alhasan, Raina, Malki, Aljamaan, Tamimi, Muaygil, Wahabi, Jamal, Al-Tawfiq, Al-Eyadhy, Soliman, and Temsah]{abouammoh_exploring_2023}
Noura Abouammoh, Khalid Alhasan, Rupesh Raina, Khalid~A. Malki, Fadi Aljamaan, Ibraheem Tamimi, Ruaim Muaygil, Hayfaa Wahabi, Amr Jamal, Jaffar~A. Al-Tawfiq, Ayman Al-Eyadhy, Mona Soliman, and Mohamad-Hani Temsah.
\newblock Exploring {Perceptions} and {Experiences} of {ChatGPT} in {Medical} {Education}: {A} {Qualitative} {Study} {Among} {Medical} {College} {Faculty} and {Students} in {Saudi} {Arabia}, July 2023.
\newblock URL \url{https://www.medrxiv.org/content/10.1101/2023.07.13.23292624v1}.
\newblock Pages: 2023.07.13.23292624.

\bibitem[Angwin et~al.(2016)Angwin, Larson, Mattu, and Kirchner]{angwin_machine_2016}
Julia Angwin, Jeff Larson, Surya Mattu, and Lauren Kirchner.
\newblock Machine {Bias}, 2016.
\newblock URL \url{https://www.propublica.org/article/machine-bias-risk-assessments-in-criminal-sentencing}.

\bibitem[Bommasani et~al.(2021)Bommasani, Hudson, Adeli, Altman, Arora, von Arx, Bernstein, Bohg, Bosselut, Brunskill, et~al.]{bommasani2021opportunities}
Rishi Bommasani, Drew~A Hudson, Ehsan Adeli, Russ Altman, Simran Arora, Sydney von Arx, Michael~S Bernstein, Jeannette Bohg, Antoine Bosselut, Emma Brunskill, et~al.
\newblock On the opportunities and risks of foundation models.
\newblock \emph{arXiv preprint arXiv:2108.07258}, 2021.

\bibitem[Bommasani et~al.(2023)Bommasani, Klyman, Longpre, Kapoor, Maslej, Xiong, Zhang, and Liang]{bommasani_foundation_2023}
Rishi Bommasani, Kevin Klyman, Shayne Longpre, Sayash Kapoor, Nestor Maslej, Betty Xiong, Daniel Zhang, and Percy Liang.
\newblock The {Foundation} {Model} {Transparency} {Index}, October 2023.
\newblock URL \url{http://arxiv.org/abs/2310.12941}.
\newblock arXiv:2310.12941 [cs].

\bibitem[Carley et~al.(2015)Carley, Hedge, and Marco]{carley2015probability}
Michael Carley, Deepak Hedge, and Alan Marco.
\newblock What is the probability of receiving a us patent.
\newblock \emph{Yale JL \& Tech.}, 17:\penalty0 203, 2015.

\bibitem[Chalkidis(2023)]{chalkidis_chatgpt_2023}
Ilias Chalkidis.
\newblock {ChatGPT} may {Pass} the {Bar} {Exam} soon, but has a {Long} {Way} to {Go} for the {LexGLUE} benchmark, March 2023.
\newblock URL \url{http://arxiv.org/abs/2304.12202}.
\newblock arXiv:2304.12202 [cs].

\bibitem[Chen et~al.(2023)Chen, Zaharia, and Zou]{chen_how_2023}
Lingjiao Chen, Matei Zaharia, and James Zou.
\newblock How is {ChatGPT}'s behavior changing over time?, October 2023.
\newblock URL \url{http://arxiv.org/abs/2307.09009}.
\newblock arXiv:2307.09009 [cs].

\bibitem[Ching and Hershkowitz(2023)]{ChingHershkowitz2023}
Audrey Ching and Donna Hershkowitz.
\newblock Report from the alternative pathway working group: Request to circulate for public comment.
\newblock Board of Trustees Meeting Agenda Item, September 2023.
\newblock URL \url{https://www.courthousenews.com/wp-content/uploads/2023/09/california-bar-exam-alternative-proposal.pdf}.
\newblock Los Angeles Office, California State Bar.

\bibitem[Corey(2019)]{corey_how_2019}
Ethan Corey.
\newblock How a {Tool} to {Help} {Judges} {May} {Be} {Leading} {Them} {Astray}, August 2019.
\newblock URL \url{https://theappeal.org/how-a-tool-to-help-judges-may-be-leading-them-astray/}.

\bibitem[Deck(2023)]{deck_ai_2023}
Andrew Deck.
\newblock {AI} translation is jeopardizing {Afghan} asylum claims, April 2023.
\newblock URL \url{https://restofworld.org/2023/ai-translation-errors-afghan-refugees-asylum/}.

\bibitem[Diver et~al.(2022)Diver, McBride, Medvedeva, Banerjee, D'hondt, Duarte~Nicolau, Dushi, Gori, Van Den~Hoven, Meessen, and Hildebrandt]{diver_typology_2022}
Laurence Diver, Pauline McBride, Masha Medvedeva, Arjun~Bhubaneshwar Banerjee, Eva D'hondt, Tatiana Duarte~Nicolau, Desara Dushi, Gianmarco Gori, Emilie Van Den~Hoven, Paulus Meessen, and Mireille Hildebrandt.
\newblock Typology of {Legal} {Technologies}: {Cross}-disciplinary {Research} in {Computational} {Law} ({CRCL}), November 2022.
\newblock URL \url{https://publications.cohubicol.com/typology/}.
\newblock Place: Brussels.

\bibitem[Dressel and Farid(2018)]{dressel_accuracy_2018}
Julia Dressel and Hany Farid.
\newblock The accuracy, fairness, and limits of predicting recidivism.
\newblock \emph{Science Advances}, 4\penalty0 (1):\penalty0 eaao5580, January 2018.
\newblock ISSN 2375-2548.
\newblock \doi{10.1126/sciadv.aao5580}.
\newblock URL \url{https://www.science.org/doi/10.1126/sciadv.aao5580}.

\bibitem[Geirhos et~al.(2020)Geirhos, Jacobsen, Michaelis, Zemel, Brendel, Bethge, and Wichmann]{geirhos_shortcut_2020}
Robert Geirhos, Jörn-Henrik Jacobsen, Claudio Michaelis, Richard Zemel, Wieland Brendel, Matthias Bethge, and Felix~A. Wichmann.
\newblock Shortcut learning in deep neural networks.
\newblock \emph{Nature Machine Intelligence}, 2\penalty0 (11):\penalty0 665--673, November 2020.
\newblock ISSN 2522-5839.
\newblock \doi{10.1038/s42256-020-00257-z}.
\newblock URL \url{https://www.nature.com/articles/s42256-020-00257-z}.
\newblock Number: 11 Publisher: Nature Publishing Group.

\bibitem[Glaze et~al.(2021)Glaze, Ho, Ray, and Tsang]{glaze2021artificial}
Kurt Glaze, Daniel~E Ho, Gerald~K Ray, and Christine Tsang.
\newblock Artificial intelligence for adjudication: The social security administration and ai governance.
\newblock 2021.

\bibitem[Gramatikov et~al.(2021)Gramatikov, Núñez, Banks, Barendrecht, Brouwer, Kauffman, and Cornett]{gramatikov_justice_2021}
Martin Gramatikov, Rodrigo Núñez, Isabella Banks, Maurits Barendrecht, Jelmer Brouwer, Brittany Kauffman, and Logan Cornett.
\newblock Justice {Needs} and {Satisfaction} in the {United} {States} of {America}, 2021.
\newblock URL \url{https://iaals.du.edu/publications/justice-needs-and-satisfaction-united-states-america}.

\bibitem[Guha et~al.(2022)Guha, Henderson, and Zambrano]{guha_vulnerabilities_2022}
Neel Guha, Peter Henderson, and Diego Zambrano.
\newblock Vulnerabilities in {Discovery} {Tech}.
\newblock \emph{Harvard Journal of Law \& Technology}, 35, 2022.
\newblock \doi{10.2139/ssrn.4065997}.
\newblock URL \url{https://jolt.law.harvard.edu/assets/articlePDFs/v35/4.-Guha-Henderson-and-Zambrano-Vulnerabilities-in-Discovery-Tech.pdf}.

\bibitem[Guha et~al.(2023)Guha, Nyarko, Ho, Ré, Chilton, Narayana, Chohlas-Wood, Peters, Waldon, Rockmore, Zambrano, Talisman, Hoque, Surani, Fagan, Sarfaty, Dickinson, Porat, Hegland, Wu, Nudell, Niklaus, Nay, Choi, Tobia, Hagan, Ma, Livermore, Rasumov-Rahe, Holzenberger, Kolt, Henderson, Rehaag, Goel, Gao, Williams, Gandhi, Zur, Iyer, and Li]{guha_legalbench_2023}
Neel Guha, Julian Nyarko, Daniel~E. Ho, Christopher Ré, Adam Chilton, Aditya Narayana, Alex Chohlas-Wood, Austin Peters, Brandon Waldon, Daniel~N. Rockmore, Diego Zambrano, Dmitry Talisman, Enam Hoque, Faiz Surani, Frank Fagan, Galit Sarfaty, Gregory~M. Dickinson, Haggai Porat, Jason Hegland, Jessica Wu, Joe Nudell, Joel Niklaus, John Nay, Jonathan~H. Choi, Kevin Tobia, Margaret Hagan, Megan Ma, Michael Livermore, Nikon Rasumov-Rahe, Nils Holzenberger, Noam Kolt, Peter Henderson, Sean Rehaag, Sharad Goel, Shang Gao, Spencer Williams, Sunny Gandhi, Tom Zur, Varun Iyer, and Zehua Li.
\newblock {LegalBench}: {A} {Collaboratively} {Built} {Benchmark} for {Measuring} {Legal} {Reasoning} in {Large} {Language} {Models}, August 2023.
\newblock URL \url{http://arxiv.org/abs/2308.11462}.
\newblock arXiv:2308.11462 [cs].

\bibitem[{Hon. Bernice Bouie Donald} et~al.(2022){Hon. Bernice Bouie Donald}, {Hon. James C. Francis IV}, {Ronald J. Hedges}, and {Kenneth J. Withers}]{hon_bernice_bouie_donald_generative_2022}
{Hon. Bernice Bouie Donald}, {Hon. James C. Francis IV}, {Ronald J. Hedges}, and {Kenneth J. Withers}.
\newblock Generative {AI} and {Courts}: {How} {Are} {They} {Getting} {Along}?, February 2022.
\newblock URL \url{https://www.jamsadr.com/blog/2023/francis-james-pli-generative-ai-1023}.

\bibitem[{Horace He [@cHHillee]}(2023)]{horace_he_chhillee_i_2023}
{Horace He [@cHHillee]}.
\newblock I suspect {GPT}-4's performance is influenced by data contamination, at least on {Codeforces}. {Of} the easiest problems on {Codeforces}, it solved 10/10 pre-2021 problems and 0/10 recent problems. {This} strongly points to contamination. 1/4 https://t.co/{wm6yP6AmGx}, March 2023.
\newblock URL \url{https://twitter.com/cHHillee/status/1635790330854526981}.

\bibitem[Jackson and Mendoza(2020)]{jackson_setting_2020}
Eugenie Jackson and Christina Mendoza.
\newblock Setting the {Record} {Straight}: {What} the {COMPAS} {Core} {Risk} and {Need} {Assessment} {Is} and {Is} {Not}.
\newblock \emph{Harvard Data Science Review}, 2\penalty0 (1), January 2020.
\newblock ISSN 2644-2353, 2688-8513.
\newblock \doi{10.1162/99608f92.1b3dadaa}.
\newblock URL \url{https://hdsr.mitpress.mit.edu/pub/hzwo7ax4/release/7}.

\bibitem[{Joshua Browder [@jbrowder1]}(2023{\natexlab{a}})]{joshua_browder_jbrowder1_donotpay_2023}
{Joshua Browder [@jbrowder1]}.
\newblock {DoNotPay} will pay any lawyer or person \$1,000,000 with an upcoming case in front of the {United} {States} {Supreme} {Court} to wear {AirPods} and let our robot lawyer argue the case by repeating exactly what it says. (1/2), January 2023{\natexlab{a}}.
\newblock URL \url{https://twitter.com/jbrowder1/status/1612312707398795264}.

\bibitem[{Joshua Browder [@jbrowder1]}(2023{\natexlab{b}})]{joshua_browder_jbrowder1_good_2023}
{Joshua Browder [@jbrowder1]}.
\newblock Good morning! {Bad} news: after receiving threats from {State} {Bar} prosecutors, it seems likely they will put me in jail for 6 months if {I} follow through with bringing a robot lawyer into a physical courtroom. {DoNotPay} is postponing our court case and sticking to consumer rights:, January 2023{\natexlab{b}}.
\newblock URL \url{https://twitter.com/jbrowder1/status/1618265395986857984}.

\bibitem[Kapoor and Narayanan(2023)]{kapoor_leakage_2023}
Sayash Kapoor and Arvind Narayanan.
\newblock Leakage and the reproducibility crisis in machine-learning-based science.
\newblock \emph{Patterns}, 4\penalty0 (9), September 2023.
\newblock ISSN 2666-3899.
\newblock \doi{10.1016/j.patter.2023.100804}.
\newblock URL \url{https://www.cell.com/patterns/abstract/S2666-3899(23)00159-9}.
\newblock Publisher: Elsevier.

\bibitem[Li and Flanigan(2023)]{li_task_2023}
Changmao Li and Jeffrey Flanigan.
\newblock Task {Contamination}: {Language} {Models} {May} {Not} {Be} {Few}-{Shot} {Anymore}, December 2023.
\newblock URL \url{https://arxiv.org/abs/2312.16337v1}.

\bibitem[Lu et~al.(2021)Lu, Bartolo, Moore, Riedel, and Stenetorp]{lu2021fantastically}
Yao Lu, Max Bartolo, Alastair Moore, Sebastian Riedel, and Pontus Stenetorp.
\newblock Fantastically ordered prompts and where to find them: Overcoming few-shot prompt order sensitivity.
\newblock \emph{arXiv preprint arXiv:2104.08786}, 2021.

\bibitem[Magar and Schwartz(2022)]{magar_data_2022}
Inbal Magar and Roy Schwartz.
\newblock Data {Contamination}: {From} {Memorization} to {Exploitation}, March 2022.
\newblock URL \url{http://arxiv.org/abs/2203.08242}.
\newblock arXiv:2203.08242 [cs].

\bibitem[Markoff(2011)]{markoff_armies_2011}
John Markoff.
\newblock Armies of {Expensive} {Lawyers}, {Replaced} by {Cheaper} {Software}.
\newblock \emph{The New York Times}, March 2011.
\newblock ISSN 0362-4331.
\newblock URL \url{https://www.nytimes.com/2011/03/05/science/05legal.html}.

\bibitem[Martínez(2023)]{martinez_re-evaluating_2023}
Eric Martínez.
\newblock Re-{Evaluating} {GPT}-4's {Bar} {Exam} {Performance}, May 2023.
\newblock URL \url{https://papers.ssrn.com/abstract=4441311}.

\bibitem[Medvedeva and Mcbride(2023)]{medvedeva_legal_2023}
Masha Medvedeva and Pauline Mcbride.
\newblock Legal {Judgment} {Prediction}: {If} {You} {Are} {Going} to {Do} {It}, {Do} {It} {Right}.
\newblock In Daniel Preo{\textbackslash}textcommabelowtiuc-Pietro, Catalina Goanta, Ilias Chalkidis, Leslie Barrett, Gerasimos~(Jerry) Spanakis, and Nikolaos Aletras, editors, \emph{Proceedings of the {Natural} {Legal} {Language} {Processing} {Workshop} 2023}, pages 73--84, Singapore, December 2023. Association for Computational Linguistics.
\newblock \doi{10.18653/v1/2023.nllp-1.9}.
\newblock URL \url{https://aclanthology.org/2023.nllp-1.9}.

\bibitem[Medvedeva et~al.(2023)Medvedeva, Wieling, and Vols]{medvedeva_rethinking_2023}
Masha Medvedeva, Martijn Wieling, and Michel Vols.
\newblock Rethinking the field of automatic prediction of court decisions.
\newblock \emph{Artificial Intelligence and Law}, 31\penalty0 (1):\penalty0 195--212, March 2023.
\newblock ISSN 1572-8382.
\newblock \doi{10.1007/s10506-021-09306-3}.
\newblock URL \url{https://doi.org/10.1007/s10506-021-09306-3}.

\bibitem[{Michael C. Dorf}(2023)]{michael_c_dorf_law-specific_2023}
{Michael C. Dorf}.
\newblock Law-{Specific} {Large} {Language} {Model} {Generative} {AI} {Interim} {Report}: {Lexis}+{AI} {Versus} {GPT}-4, November 2023.
\newblock URL \url{https://www.dorfonlaw.org/2023/11/law-specific-large-language-model.html}.

\bibitem[Moore(2017)]{moore_trade_2017}
Taylor Moore.
\newblock Trade {Secrets} and {Algorithms} as {Barriers} to {Social} {Justice}, August 2017.
\newblock URL \url{https://cdt.org/insights/trade-secrets-and-algorithms-as-barriers-to-social-justice/}.

\bibitem[Narayanan and Kapoor(2023{\natexlab{a}})]{narayanan_generative_2023}
Arvind Narayanan and Sayash Kapoor.
\newblock Generative {AI} companies must publish transparency reports, 2023{\natexlab{a}}.
\newblock URL \url{http://knightcolumbia.org/blog/generative-ai-companies-must-publish-transparency-reports}.

\bibitem[Narayanan and Kapoor(2023{\natexlab{b}})]{narayanan_is_2023}
Arvind Narayanan and Sayash Kapoor.
\newblock Is {GPT}-4 getting worse over time?, July 2023{\natexlab{b}}.
\newblock URL \url{https://www.aisnakeoil.com/p/is-gpt-4-getting-worse-over-time}.

\bibitem[Nayak et~al.(2023)Nayak, Alkaitis, Nayak, Nikolov, Weinfurt, and Schulman]{nayak_comparison_2023}
Ashwin Nayak, Matthew~S. Alkaitis, Kristen Nayak, Margaret Nikolov, Kevin~P. Weinfurt, and Kevin Schulman.
\newblock Comparison of {History} of {Present} {Illness} {Summaries} {Generated} by a {Chatbot} and {Senior} {Internal} {Medicine} {Residents}.
\newblock \emph{JAMA Internal Medicine}, 183\penalty0 (9):\penalty0 1026--1027, September 2023.
\newblock ISSN 2168-6106.
\newblock \doi{10.1001/jamainternmed.2023.2561}.
\newblock URL \url{https://doi.org/10.1001/jamainternmed.2023.2561}.

\bibitem[Noy and Zhang(2023)]{noy_experimental_2023}
Shakked Noy and Whitney Zhang.
\newblock Experimental evidence on the productivity effects of generative artificial intelligence.
\newblock \emph{Science}, 381\penalty0 (6654):\penalty0 187--192, July 2023.
\newblock \doi{10.1126/science.adh2586}.
\newblock URL \url{https://www.science.org/doi/10.1126/science.adh2586}.
\newblock Publisher: American Association for the Advancement of Science.

\bibitem[OpenAI(2023)]{openai_gpt-4_2023}
OpenAI.
\newblock {GPT}-4 {Technical} {Report}, March 2023.
\newblock URL \url{http://arxiv.org/abs/2303.08774}.
\newblock arXiv:2303.08774 [cs].

\bibitem[{Paris Martineau}(2022)]{paris_martineau_toronto_2022}
{Paris Martineau}.
\newblock Toronto {Tapped} {Artificial} {Intelligence} to {Warn} {Swimmers}. {The} {Experiment} {Failed}, 2022.
\newblock URL \url{https://www.theinformation.com/articles/when-artificial-intelligence-isnt-smarter}.

\bibitem[Raji et~al.(2021)Raji, Denton, Bender, Hanna, and Paullada]{raji_ai_2021}
Deborah Raji, Emily Denton, Emily~M. Bender, Alex Hanna, and Amandalynne Paullada.
\newblock {AI} and the {Everything} in the {Whole} {Wide} {World} {Benchmark}.
\newblock \emph{Proceedings of the Neural Information Processing Systems Track on Datasets and Benchmarks}, 1, December 2021.
\newblock URL \url{https://datasets-benchmarks-proceedings.neurips.cc/paper/2021/hash/084b6fbb10729ed4da8c3d3f5a3ae7c9-Abstract-round2.html}.

\bibitem[Shuster et~al.(2021)Shuster, Poff, Chen, Kiela, and Weston]{shuster_retrieval_2021}
Kurt Shuster, Spencer Poff, Moya Chen, Douwe Kiela, and Jason Weston.
\newblock Retrieval {Augmentation} {Reduces} {Hallucination} in {Conversation}.
\newblock In Marie-Francine Moens, Xuanjing Huang, Lucia Specia, and Scott Wen-tau Yih, editors, \emph{Findings of the {Association} for {Computational} {Linguistics}: {EMNLP} 2021}, pages 3784--3803, Punta Cana, Dominican Republic, November 2021. Association for Computational Linguistics.
\newblock \doi{10.18653/v1/2021.findings-emnlp.320}.
\newblock URL \url{https://aclanthology.org/2021.findings-emnlp.320}.

\bibitem[Sloan(2023)]{Sloan2023}
Karen Sloan.
\newblock New bar exam gets lukewarm reception in previews, 2023.
\newblock URL \url{https://www.reuters.com/legal/legalindustry/new-bar-exam-gets-lukewarm-reception-previews-2023-07-19/}.
\newblock Accessed: 2023-11-08.

\bibitem[{Stanford Legal Design Lab} and {Suffolk LIT Lab}(2018)]{stanford_legal_design_lab_learned_2018}
{Stanford Legal Design Lab} and {Suffolk LIT Lab}.
\newblock Learned {Hands}, 2018.
\newblock URL \url{https://learnedhands.law.stanford.edu}.

\bibitem[{Statista Research Department}(2023)]{statista_research_department_us_2023}
{Statista Research Department}.
\newblock U.{S}.: number of lawyers 2007-2022, 2023.
\newblock URL \url{https://www.statista.com/statistics/740222/number-of-lawyers-us/}.

\bibitem[Thomas and Uminsky(2022)]{thomas_reliance_2022}
Rachel~L. Thomas and David Uminsky.
\newblock Reliance on metrics is a fundamental challenge for {AI}.
\newblock \emph{Patterns}, 3\penalty0 (5):\penalty0 100476, May 2022.
\newblock ISSN 2666-3899.
\newblock \doi{10.1016/j.patter.2022.100476}.
\newblock URL \url{https://www.sciencedirect.com/science/article/pii/S2666389922000563}.

\bibitem[Vincent(2023)]{vincent_openai_2023}
James Vincent.
\newblock {OpenAI} isn’t doing enough to make {ChatGPT}’s limitations clear, May 2023.
\newblock URL \url{https://www.theverge.com/2023/5/30/23741996/openai-chatgpt-false-information-misinformation-responsibility}.

\bibitem[Wagner(2023)]{wagner_this_2023}
David Wagner.
\newblock This {Prolific} {LA} {Eviction} {Law} {Firm} {Was} {Caught} {Faking} {Cases} {In} {Court}. {Did} {They} {Misuse} {AI}?, October 2023.
\newblock URL \url{https://laist.com/news/housing-homelessness/dennis-block-chatgpt-artificial-intelligence-ai-eviction-court-los-angeles-lawyer-sanction-housing-tenant-landlord}.
\newblock Section: Housing and Homelessness.

\bibitem[Wang et~al.(2023)Wang, Kapoor, Barocas, and Narayanan]{wang_against_2023}
Angelina Wang, Sayash Kapoor, Solon Barocas, and Arvind Narayanan.
\newblock Against {Predictive} {Optimization}: {On} the {Legitimacy} of {Decision}-{Making} {Algorithms} that {Optimize} {Predictive} {Accuracy}.
\newblock In \emph{Proceedings of the 2023 {ACM} {Conference} on {Fairness}, {Accountability}, and {Transparency}}, {FAccT} '23, page 626, New York, NY, USA, June 2023. Association for Computing Machinery.
\newblock ISBN 9798400701924.
\newblock \doi{10.1145/3593013.3594030}.
\newblock URL \url{https://dl.acm.org/doi/10.1145/3593013.3594030}.

\bibitem[Weiser(2023)]{weiser_heres_2023}
Benjamin Weiser.
\newblock Here’s {What} {Happens} {When} {Your} {Lawyer} {Uses} {ChatGPT}.
\newblock \emph{The New York Times}, May 2023.
\newblock ISSN 0362-4331.
\newblock URL \url{https://www.nytimes.com/2023/05/27/nyregion/avianca-airline-lawsuit-chatgpt.html}.

\bibitem[Zhao et~al.(2020)Zhao, Cohen, and Webber]{zhao_reducing_2020}
Zheng Zhao, Shay~B. Cohen, and Bonnie Webber.
\newblock Reducing {Quantity} {Hallucinations} in {Abstractive} {Summarization}.
\newblock In Trevor Cohn, Yulan He, and Yang Liu, editors, \emph{Findings of the {Association} for {Computational} {Linguistics}: {EMNLP} 2020}, pages 2237--2249, Online, November 2020. Association for Computational Linguistics.
\newblock \doi{10.18653/v1/2020.findings-emnlp.203}.
\newblock URL \url{https://aclanthology.org/2020.findings-emnlp.203}.

\bibitem[Zheng et~al.(2023)Zheng, Chiang, Sheng, Li, Zhuang, Wu, Zhuang, Li, Lin, Xing, Gonzalez, Stoica, and Zhang]{zheng_lmsys-chat-1m_2023}
Lianmin Zheng, Wei-Lin Chiang, Ying Sheng, Tianle Li, Siyuan Zhuang, Zhanghao Wu, Yonghao Zhuang, Zhuohan Li, Zi~Lin, Eric~P. Xing, Joseph~E. Gonzalez, Ion Stoica, and Hao Zhang.
\newblock {LMSYS}-{Chat}-{1M}: {A} {Large}-{Scale} {Real}-{World} {LLM} {Conversation} {Dataset}, September 2023.
\newblock URL \url{http://arxiv.org/abs/2309.11998}.
\newblock arXiv:2309.11998 [cs].

\end{thebibliography}


\end{document}